\documentclass[12pt]{iopart}
\usepackage{epsfig}
\usepackage{cite} 
\usepackage{url}  
\usepackage{amssymb}
                                                                                
\begin{document}

\title{The cytoplasm of living cells: A functional mixture of
thousands of components}
\author{Richard P. Sear}
                                                                                
\address{
Department of Physics, University of Surrey,\\
Guildford, Surrey GU2 7XH,
United Kingdom\\
email: r.sear@surrey.ac.uk}

\begin{abstract}
Inside every living cell is the cytoplasm: a fluid mixture
of thousands of different macromolecules,
predominantly proteins. This mixture is where most of the
biochemistry occurs that enables living cells to function, and
it is perhaps the most complex liquid on earth. Here
we take an inventory of what is actually in this mixture.
Recent genome-sequencing work has given us for the first time
at least some information on all of these thousands of components.
Having done so we consider two physical phenomena in the
cytoplasm: diffusion and possible phase separation.
Diffusion is slower in the highly crowded cytoplasm
than in dilute solution.
Reasonable estimates of this slowdown can be obtained and their
consequences explored, for example, monomer-dimer equilibria
are established approximately twenty times slower than in
a dilute solution. Phase separation in all except exceptional cells
appears not to be a problem, despite the high density and so
strong protein-protein interactions present. We suggest that
this may be partially a byproduct of the evolution of other
properties, and partially a result of the huge number of components
present.
\end{abstract}

\maketitle

\section{Introduction to the cytoplasm}

Living cells are essentially
very complex membranes surrounding equally complex solutions
of, predominantly, protein molecules. These solutions are
arguably the most complex liquids we know of. This article
will begin with some of the basic questions we can ask about these
complex liquids, together with some partial answers.
Then we will look at
two phenomena in the cytoplasm that are particularly suited to
study by physical scientists:
diffusion and phase behaviour. Below there is a section on each,
and we will end with a brief conclusion.
In the following section we will look at two aspects of diffusion in the
crowded environment of the cell.
The first is the need to
estimate the slow down due to the high density of protein present. The
second aspect is cytoplasmic diffusion as a process that
has been optimised by evolution. If say the rate of diffusion
is limiting the speed of response of the cell to a change in the
environment then there is natural selection pressure on the
proteins to evolve to diffuse faster.
Section \ref{sec_phasesep} will discuss how we can understand
and even calculate some aspects
of the phase behaviour of models of the cytoplasm,
even in the absence of hard data on even one of the millions
of interactions that occur in the cytoplasm.
In the remainder of this introduction we will
consider some of the basic questions we can ask about the cytoplasm.

{\em What is in it?} A concentrated solution of macromolecules,
predominantly protein, but also RNA and in the case of prokaryotes
one or a few huge DNA molecules. Proteins are heteropolymers,
they are linear chains
of amino acids that are typically folded up into a compact,
relatively rigid native state that is in many ways more
like a colloid than a conventional polymer.
Prokaryote cells are much
simpler than those of eukaryotes. Prokaryotes are
(relatively) simple organisms such as bacteria, e.g., {\em E.~coli}.
Their cells contain only one compartment that contains the DNA,
the proteins, the ribosomes where new proteins are made etc.
See any molecular biology textbook, for example
that of Alberts {\it et al.} \cite{thecell}.
Eukaryote cells are larger and compartmentalised, in particular
the DNA is in a membrane-bound compartment called the nucleus, not
in the cytoplasm. Some eukaryotes are single-celled organisms,
e.g., yeast, but all complex multicellular organisms,
e.g., {\em H.~sapiens}, are eukaryotes.
Eukaryote cells have a complex \lq skeleton' of filaments of protein
\cite{thecell}
and not all of the protein diffuses freely in the cytoplasm
\cite{pagliaro88}. We will not discuss this further here but
it should be borne in mind that the description of the
cytoplasm as a liquid mixture may be a better approximation
in prokaryotes than in eukaryotes. See \cite{lubyphelps00}
for a review of the properties of the eukaryote cytoplasm.

Returning to prokaryotes, an inventory
of the protein, RNA and DNA in {\em E.~coli} is given in Table \ref{t1}.
For their net electrostatic charges see \cite{sear03prot}.
The bacterium {\it E.~coli} has been extensively studied
and much is known about it \cite{neidhardt,coligen,ebi}.
The macromolecules occupy around 30\% to 40\% of the volume inside
the cell. Of course it is well known that at these concentrations the
interactions between the molecules are both strong and important.

\begin{table}[t]
\begin{center}
\begin{tabular}{|c|c|c|c|c|}
\hline
 & volume (nm$^3$) & no.~of types & no. of molecules & volume fraction \\
\hline
Protein & 100 &  1000 & $10^6$ & 10\%  \\
tRNA & 100 &  10 & $10^5$ & 1\% \\
Ribosome & $10^4$ & 1 & $10^4$ & 10\% \\
DNA & $10^6$ & 1  & 1 & $0.1$\% \\
\hline
\end{tabular}
\caption{The protein, RNA and DNA in the cytosol of {\em E. coli}
\cite{neidhardt,coligen,ebi,sear03prot}.
For each class of macromolecule,
the columns indicate the orders of magnitude
of, from left to right:
the volume of a single molecule of this class,
the number of different types of molecule in this class,
the total number of molecules
in the cytoplasm of a cell, and the
volume fraction occupied by molecules of this
class. The cytoplasm has a volume of order
1$\mu$m$^3$. Ribosomes are large complexes of protein and RNA. They are
the cell's protein factories. tRNA molecules are
relatively small RNA molecules that hold an amino acid
in readiness for it to be added to the growing chain
of amino acids that is being synthesised at a ribosome.
\label{t1}
}
\end{center}
\end{table}

{\em What does it do?} Living organisms consume energy,
grow, move etc. The cytoplasm is where most of the energy
is consumed and most of the functions necessary
to grow etc are performed. The cytoplasm also computes: it receives and
integrates signals from the environment and changes the
functions performed accordingly.
For example, if the environment of
{\em E.~coli} contains the sugar lactose but not glucose then a signal
is transmitted within the cytoplasm and the synthesis of
the enzymes needed to metabolise this sugar is
switched on \cite{thecell}.

{\em How does it compute responses, copy DNA etc?}
This is a large question, indeed
essentially all of cell biology is concerned with answering this
question. Liquids physicists perhaps have most to contribute
to processes that either involve transport, such as diffusion,
or the underlying equilibrium behaviour of the mixture
of complex molecules that forms the cytoplasm. Thus we will
focus on diffusion in section \ref{sec_diff}
and phase behaviour in section \ref{sec_phasesep}.

\section{Diffusion {\it in vivo}}
\label{sec_diff}

As a first example, let us consider diffusion. This
is essential to transmit signals across the cytoplasm, for
reactants to collide and so on. We want to
understand diffusion {\it in vivo}, i.e., in the cytoplasm,
and to do so we will compare diffusion {\it in vivo}
with diffusion {\it in vitro}, by which we mean diffusion in the
typically very dilute solutions that
biochemists study. These solutions are so dilute that
they can be treated as an ideal gas.

The properties of the cytoplasm
have been optimised by almost four billion
years of evolution. Thus one approach to understanding the
cytoplasm is to consider how it can be optimised.
See for example the work of Bialek \cite{bialek}
for elegant examples of this approach.
If we consider diffusion-limited reactions between a pair of
proteins $A$ and $B$, then the reaction rate is \cite{atkins}
\begin{equation}
\mbox{Rate} = k N_AN_B/V_{CYTO},
\label{cytoreac}
\end{equation}
for $N_A$ molecules of protein $A$, and $N_B$ molecules of protein
$B$ uniformly
distributed within a cytoplasm of volume $V_{CYTO}$. The reaction constant
$k\approx Dr$, where $D$ and $r$ are the diffusion constant and the
linear dimension of the volume within which the reaction
occurs, respectively \cite{atkins}.
Thus the reaction rate per $N_A$ molecule
is proportional to $kN_B/V_{CYTO}$.

For the sake of argument, let us guess that the total volume
of the cytoplasm $V_{CYTO}$ is
is determined by the need to maximise reaction rates
such as that of equation (\ref{cytoreac}).
Bacteria are under strong
natural selection pressure to be able to grow rapidly and if
reactions like that of equation (\ref{cytoreac}) limit this
rate then there will be selection pressure to speed up the reaction.
At fixed numbers of proteins varying the volume
fraction of protein $\phi$ is equivalent to varying the volume $V_{CYTO}$.
Thus, we will search for the value of $\phi$ that maximises
reaction rates.
Now, the reaction rate per $A$ molecule depends
on $\phi$ in two ways: i) $N_B/V_{CYTO}\propto \phi$,
the denser the cytoplasm the higher the density of $B$ molecules, and
ii) through the reaction constant $k=D(\phi)r$, which depends
on the density-dependent diffusion constant.

Thus, the reaction rate per $A$ molecule is proportional
to $D(\phi)\phi$ and
{\em if} the density of the cytoplasm is set by
the requirement to maximise the reaction rate of diffusion-limited
reactions then we expect to find cells with a volume fraction
$\phi$ that maximises $D(\phi)\phi$. Clearly, many other things
are going on in a cell that need to be optimised other than the
rate of diffusion but let us persevere with our naive assumption.
We do not know the density dependence of the self-diffusion
constant in the cytoplasm, although Elowitz {\it et al.}
\cite{elowitz99} have measured the diffusion constant
for a small protein in {\em E.~coli}. So, we resort to the
standard, but drastic physicists' approximation of treating
proteins as hard spheres. A reasonable
approximation to the long-time self-diffusion constant
of colloidal hard spheres is given by
\cite{blaaderen92}
\begin{equation}
D(\phi)=D_0\frac{(1-\phi)^3}{1+(3/2)\phi+2\phi^2+3\phi^3},
\label{diff}
\end{equation}
where $D_0=kT/6\pi\eta a$ is the Stokes-Einstein expression
for the diffusion constant for a colloidal particle at
infinite dilution. $kT$ is the thermal energy, $\eta$ is the
viscosity of the solvent, here a salt solution, and $a$ is the
radius of the colloidal particle. Equation (\ref{diff}) is based
on earlier work by Medina-Noyola \cite{medina-noyola88}.
Using equation (\ref{diff}) for $D(\phi)$, we find that the
$D(\phi)\phi$ is maximal at $\phi=0.18$, rather lower than
that found inside cells. Also, at this volume fraction the
self-diffusion constant is $40\%$ of its value at infinite dilution
whereas the measurements of Elowitz {\it et al.} \cite{elowitz99}
put the diffusion constant of a protein
called Green Fluorescent Protein (GFP), in {\it E.~coli}
at approximately $10\%$
of its value in a dilute solution. Thus our very
naive assumption that
the cytoplasm is effectively a hard-sphere suspension optimised
for the reaction rate between
pairs of proteins is not consistent with the experimental
data. However, note that equation (\ref{diff}) predicts that at
volume fractions $\phi=0.3$ and $0.4$ the self-diffusion constant
is $0.2$ and $0.1$ times its value at infinite dilution, respectively.
Thus,
given the density of the cytoplasm the speed of diffusion
in the cytoplasm,
at least of some relatively small proteins, is similar to that
in a hard-sphere suspension with the same volume fraction.
In summary, it is possible that the proteins are mostly
not very sticky and so on average the interactions are
not far from simple hard-sphere-like repulsions, but the
density of the cytoplasm appears to be too high to be the
result of selection for the maximum collision rate between
proteins.

\begin{figure}[t]
\vspace*{1.0cm}
\begin{center}
\epsfig{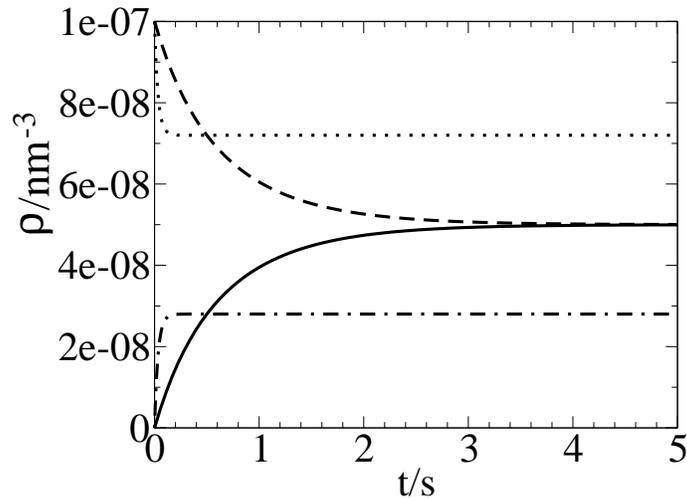}
\end{center}
\caption{{
A plot of the concentrations of both the monomers
of type $A$ and the $AB$ dimers, as a function of time.
The dashed and solid curves are $\rho_A$ and $\rho_{AB}$,
respectively, for the {\it in vivo} reaction.
The dotted and dot-dashed curves are $\rho_A$ and $\rho_{AB}$,
respectively, for the {\it in vitro} reaction.
The parameter values are as described in the text, and
in each case the initial concentrations were
$\rho_A=\rho_B=10^{-7}$nm$^{-3}$, and
$\rho_{AB}=0$. As their starting concentrations were the same
the densities of the $B$ proteins were at all times equal to the
densities of the $A$ proteins and so we have not plotted the
densities of the $B$ proteins.
\label{figreac}
}}
\end{figure}

The dense protein solution that is the {\it in vivo}
environment will affect not only the rate at which a
pair of proteins come together but also the rate at which
a dimer will break apart --- we expect the push of the other molecules
will make it harder for the proteins of a dimer to move away from
each other. To study this effect let us consider that when proteins
$A$ and $B$ collide and react, with a rate given by equation
(\ref{cytoreac}), they form a dimer that then persists
for some time before dissociating. Then there will be an
an equilibrium between $A$ and $B$ monomers
and $AB$ dimers,
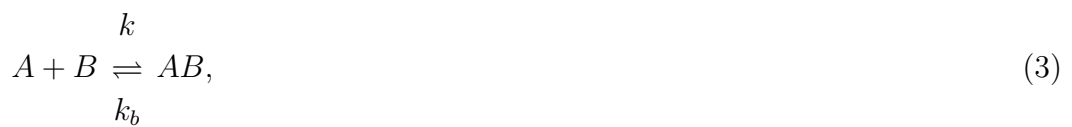
\begin{equation}
A+B
\begin{array}{c}
k\\
\rightleftharpoons\\
k_b
\end{array}
AB,
\end{equation}
with forward and back rate constants $k$ and $k_b$, respectively.
We will study the effect of the {\it in vivo} environment
by comparing dimer formation there with dimer formation in
a dilute solution {\it in vitro}. The {\it in vitro} situation
is taken to be a typical experimental situation where the proteins
are so dilute that they are an associating ideal gas. See
\cite{sear95,zhou92} for the theory of associating ideal gases
and for association in dense liquids.

Now, let us consider proteins $A$ and $B$ that each exist as 100 copies in a
cytoplasm of volume
$V_{CYTO}=1\mu$m$^3$. Thus the total number densities
of $A$ and $B$, $\rho_A+\rho_{AB}=\rho_B+\rho_{AB}=10^{-7}$nm$^{-3}$.
$\rho_A=N_A/V_{CYTO}$, $\rho_B=N_B/V_{CYTO}$ and $\rho_{AB}=N_{AB}/V_{CYTO}$,
where $N_{AB}$ is the number of $AB$ dimers in the cytoplasm.
In order to have comparable amounts of the dimer and the
free $A$ and $B$ proteins {\em in vivo} we set the dissociation
constant in the cytoplasm $K_d^c=5\times10^{-8}$nm$^{-3}$. Note that
the dissociation constant is, by definition, one over the
equilibrium constant, $K_d=\rho_A\rho_B/\rho_{AB}$.
Assuming as above that the interactions
can be modelled by hard core interactions, and
setting the hard-sphere
volume fraction $\phi=0.4$, then the dissociation constant
{\it in vitro} is given by \cite{zhou92,sear95}
\begin{equation}
K_d^v=K_d^cg_{HS}(\phi=0.4),
\label{kdr}
\end{equation}
where $g_{HS}$ is the pair distribution function at contact.
Using the Carnahan-Starling \cite{carnahan69} equation,
$g_{HS} (\phi=0.4)=3.70$. Then we obtain $K_d^v=1.85\times 10^{-7}$nm$^{-3}$.
We are assuming that the dimer consists of a pair of touching
hard spheres, as in \cite{sear95}.
Equation (\ref{kdr}) follows directly from
the definition of the pair distribution
function. The pair distribution function is the ratio of
the actual probability of finding a pair at a given separation
to the probability of finding a pair at that
separation in the absence of interactions,
i.e., in an ideal gas. See \cite{zhou92,sear95} for the use
of pair distribution functions in obtaining the density
dependence of monomer-dimer equilibria.

We use $k=D_0r$ with $r=1$nm for the rate constant
for the forward reaction {\it in vitro}
and $k=D(\phi=0.4)r$ {\it in vivo}.
Taking $T=298$K and $\eta=10^{-3}$Pa~s for water,
then using the Stokes-Einstein expression
the diffusion constant for
a protein with diameter 5nm is $D_0=87\mu$m$^2$s$^{-1}$.
In the cytoplasm, $D(\phi=0.4)=8.9\mu$m$^2$s$^{-1}$.
These values are similar to those for the protein GFP, whose
{\it in vitro} and {\it in vivo} diffusion constants are
$87\mu$m$^2$s$^{-1}$ \cite{elowitz99,swaminathan97} and
$7.7\mu$m$^2$s$^{-1}$ \cite{elowitz99}, respectively.
As $K_d=k_b/k$, then once we have specified both the rate
constant for the forward reaction and the dissociation constant
we have the rate constant for the back reaction \cite{atkins}.
Here the rate constants for the back reaction, $k_b$, are
$0.45$s$^{-1}$ {\it in vivo} and $16$s$^{-1}$ {\it in vitro}.

It is straightforward to obtain the concentrations $\rho_A$,
$\rho_B$ and $\rho_{AB}$ as functions of time {\it in vivo} and
{\it in vitro}, by in each case solving the equations
\begin{eqnarray}
\frac{{\rm d}\rho_{\alpha}}{{\rm d}t} & =&
-k\rho_A\rho_B+k_b\rho_{AB}~~~~\alpha=A,B
\label{ode1}\\
\frac{{\rm d}\rho_{AB}}{{\rm d}t} & =&
k\rho_A\rho_B-k_b\rho_{AB}.
\label{ode2}
\end{eqnarray}
after setting the initial conditions.
We use the initial condition that the density
of $AB$ dimers is zero. The solutions to
equations (\ref{ode1}) and (\ref{ode2}) are plotted in figure
\ref{figreac}. The reaction in the cytoplasm
takes around 4s to reach an equilibrium of equal numbers
of monomers and dimers, while the reaction in dilute solution takes
around $0.2$s to reach an equilibrium in which there are two and
a half times
as many monomers as dimers. Thus although the behaviour
is qualitatively the same in both cases, quantitatively there
is a large difference. Of course, we have assumed that the
interactions are hard-sphere like, attractions will alter the
picture.

The dissociation constant {\it in vitro} is a factor
of $g_{HS}(\phi=0.4)=3.70$ larger than {\it in vivo}, which means
that taking the proteins $A$ and $B$ out of the {\it in vivo}
environment and putting them in a dilute solution will significantly
reduce the number of dimers formed. The factor of $3.70$ is for a
pair of proteins $A$ and $B$ of sizes comparable to the average size of the
proteins whose crowding is pushing them together. The
dissociation constant {\it in vitro} will be increased by larger factors
if the two species are larger than typical proteins.
Thus, processes that involve the assembly of large complexes
should be especially strongly affected by removal from the
{\it in vivo} environment. The copying of DNA is one such process,
it is done by the cooperative action of a complex of a number of
proteins that bind to the double helix of DNA.
Studies of
the copying of DNA \cite{kornberg00,zimmerman87} found that
it was impossible to initiate the
copying process {\it in vitro} unless a concentrated
solution of the water-soluble polymer polyethylene glycol
was added to compensate
for the lack of the concentrated solution of other proteins found {\it in vivo}
\cite{kornberg00}. This prompted Kornberg to make compensating
in {\it in vitro} systems for the effect of taking the system
being studied out of its {\it in vivo} home one of his
ten commandments of biochemistry.
Minton has written
or cowritten a number
of reviews on the effect of the crowded
environment {\it in vivo} \cite{minton01,hall03}.

\section{Phase separation in the cytoplasm}
\label{sec_phasesep}

The cytoplasm of all but exceptional cells seems to be highly stable
with respect to phase transitions such as demixing or crystallisation.
Bacteria such as {\em E.~coli} can survive rather large
changes in the physical properties of their environment, such
as its osmotic pressure \cite{record98}, without the cytoplasm becoming
thermodynamically unstable. Phase-transition phenomena
have been observed in the cells that make up the lens of the eye
\cite{pande01} but these cells are exceptional. They are
inert and the cytoplasm is predominantly composed of families
of proteins called the $\alpha$, $\beta$ and $\gamma$-crystallins
\cite{siezen85}. This is very different from the composition of Table \ref{t1}.
Phase transitions in lens cells have been studied extensively
as they are implicated in the formation of cataracts.

We do not know why phase separation occurs
in exceptional cells such as those in the lens
of the eye but does not seem to occur in prokaryote cells
or \lq normal' human cells.
However, here we will briefly consider a speculative explanation
for the lack of attractive protein-protein interactions that
could cause separation into protein-rich and protein-poor
phases, and also a possible explanation
for the lack of demixing into phases with different protein
compositions. Note also that the selection pressure acts
on proteins in their natural habitat: the cytoplasm, but affects
their {\it in vitro} properties. For example,
Doye {\it et al.} \cite{doye04} have
speculated that proteins may be under significant selection pressure
not to crystallise and that this may contribute to the
difficulty protein crystallographers have in crystallising
proteins.

Our first speculation is that the stability of the cytoplasm is
a byproduct of selection for another property.
This property may be diffusion in the cytoplasm.
We have already considered
diffusion and the rates of diffusion-limited reactions and found
that at least for some small proteins the measured diffusion
\cite{elowitz99} is consistent with hard-sphere-like interactions.
Clearly, if many protein-protein interactions have been selected
to be hard-sphere-like in order to speed the diffusion
of the protein molecules then this
will as a byproduct select against separation into
protein-rich and protein-poor phases as the attractions
required for this form of phase transition will be selected against.

The second speculation is that a demixing phase transition, i.e.,
phase separation into two phases with similar total protein
concentrations but different compositions,
is suppressed by the central-limit theorem of statistics.
Phase transitions are driven by interactions and so to understand
how this might come about we need to consider the effect
of interactions on the thermodynamic functions of the mixture.
At the simplest level the interactions affect these functions
via the second virial coefficients. At the
second-virial-coefficient level the excess chemical potential
of component $i$ of an $N$ component mixture is
\begin{equation}
\mu_{Xi}=2\sum_{j=1}^NB_{ij}\rho_j,
\label{muxi}
\end{equation}
where $B_{ij}$ is the second virial coefficient for the interaction
between components $i$ and $j$, and $\rho_j$ is the number density
of component $j$.

In earlier work Cuesta and the author
assumed that the second virial coefficients $B_{ij}$ were independent
random variables \cite{sear03}.
Their arguments for taking this apparently rather radical step were as
follows. Firstly, the cytoplasm may contain thousands of proteins and hence
even at the level of the second-virial coefficients requires
millions of coefficients to describe the interactions. We know none
of these virial coefficients and so have no choice but to guess them.
Secondly, work by nuclear physicists on the spectra of complex
nuclei has shown that by guessing the elements of the Hamiltonian
matrix of the nuclei, some experimental observations can be reproduced, see
for example \cite{wigner67}. Inspired by this work Sear and Cuesta
replaced the matrix of second virial coefficients by a random matrix.

Having assumed not only that the $B_{ij}$ are random variables
but that they are independent, we can easily obtain the
probability distribution function of the
excess chemical potential $\mu_X$.
We denote the mean and standard deviation of the $B_{ij}$ by
$b$ and $\sigma$, respectively. Then
the central limit theorem tells us \cite{degroot}
that in the large $N$ limit,
the probability distribution function for the excess
chemical potential of a component in the mixture is the Gaussian
\begin{equation}
p\left(\mu_{X}\right)=\frac{1}{\left(2\pi\sigma_X^2\right)^{1/2}}
\exp\left[-\left(\mu_{X}-{\overline\mu_X}\right)^2/
\left(2\sigma_X^2\right)\right]
\label{pdf}
\end{equation}
with mean ${\overline\mu_X}=2b\sum_{i}^N\rho_j=2b\rho_T$,
where $\rho_T$ is the total density, and a standard
deviation given by $\sigma_X^2=4\sigma^2\sum_{i}^N\rho_j^2$.
Note that once the second virial coefficients are assumed
to be random variables, the excess chemical potential
is also a random variable,
and so a probability distribution function is the appropriate
description.
As the number of components $N$ increases at fixed total
density $\rho_T$, the individual densities must scale as $1/N$,
and so the variance $\sigma_X^2$ will also scale as $1/N$ ---
it tends to zero. Thus, as the width of the probability
distribution for the excess chemical potentials is tending
to zero as $N\to\infty$, the excess chemical potentials of
all components tend to the same value. Then, the effect
of the interactions on the chemical potentials of all
components are the same and so the mixture behaves as a
single component system in so far as the interactions are concerned.
This of course rules out demixing into phases that have the
same total concentration of protein but different compositions.

In addition, if indeed the $B_{ij}$ can be modelled by random variables
and the correlations between them are weak it suggests that
the differences between the excess chemical potentials of
a given protein in the cytoplasms of different prokaryotes,
for example in {\em E.~coli} and in {\em M.~tuberculosis},
may be small. Although
the proteins in the different species may differ in many ways, the
sum of the effects of these differences is small as the individual
effects tend to average out.

In the previous section, we noted that the volume fraction
of macromolecules in the cytoplasm was around $\phi=0.4$, which
is too high for a virial expansion truncated after the
second-virial coefficient terms to be a good description of the
free energy. However, our result that the excess chemical potentials
of the components should become increasingly similar as $N$
increases is not restricted to the second-virial-coefficient
approximation. Let us consider the situation where
the excess chemical potential of the $i$th
component is given not by equation (\ref{muxi}) but by the more general
expression
\begin{equation}
\mu_{Xi}=f(\phi)+\sum_{j=1}^N\chi_{ij}\rho_j
+\sum_{j,k=1}^N\chi_{ijk}\rho_j\rho_k+\cdots
\label{muxi2}
\end{equation}
where the $\chi_{ij}$ and $\chi_{ijk}$ control the component-dependent
contributions to the excess chemical potential and
are independent random variables like the $B_{ij}$, and $f(\phi)$ is
some function of volume fraction which accounts for component-independent
contributions such as excluded volume. For large $N$,
equation (\ref{muxi2}) gives a probability-distribution function
for the excess chemical potential with the same form as equation
(\ref{pdf}), and with a standard deviation whose largest term
for large $N$ also scales as $1/N^{1/2}$. Thus, a whole
class of mixtures in which the interactions between components
$i$ and $j$ can be modelled by independent random variables,
behave as single-component mixtures in the large $N$ limit.
This finding is in no way restricted to the small volume fractions
at which the second-virial-coefficient approximation is accurate.

The above arguments against demixing are only indicative, a careful
analysis of demixing in mixtures with virial coefficients that
are independent random variables is in
\cite{sear03}. The arguments rely on the assumption that the $B_{ij}$
are {\em independent} random variables. Correlations between
a specific property of a protein, namely its size, and its
interactions were considered by Braun {\it et al.} \cite{braun05}
using the theory of polydisperse mixtures (reviewed in \cite{sollich02}).
They considered, as a model of the mixture of proteins inside
cells, a mixture of spherical particles with a \lq stickiness' between
proteins with $l_i$ and $l_j$ amino acids that scales
as $l^{2/3}_i+l^{2/3}_j$, i.e., the contribution
of the stickiness to the second virial coefficient scaled with the
sum of the surface areas of the interacting proteins. This
is reasonable if the surfaces of proteins differ weakly
from one protein to another, and the attractive interaction
when these surfaces approach each other
is indeed a \lq sticky' interaction, i.e., has
a range that is small in comparison to the diameter of the proteins.
Braun {\it et al.}
used both genome data for the numbers of amino acids in all the
proteins for a number of organisms, and experimental proteomics
data for a bacterium that infects salmon, and found
that with this model the width of the distribution in virial coefficients due
to the distribution of protein lengths was far too small
to induce demixing \cite{braun05}. Proteomics is the study of
the complete set of proteins of an organism. The finding that,
for a simple model, the systematic variation of protein-protein
interactions with a property of the proteins, here size, has little
effect, broadly supports the use of uncorrelated random variables
to represent the virial coefficients.

\section{Conclusion}

The solutions inside living cells are perhaps the most complex liquids
on earth. They contain thousands of complex components and they
are non-equilibrium systems. This complexity is daunting but
as we have seen in the previous section, a statistical
approach can be used to make progress. Such an approach
can be used to model the effect of the interactions
in the crowded cytoplasm on any property, for example
the rate of protein unfolding \cite{sear04unf}.

The very high density of macromolecules in the cytoplasm means
that virtually all processes that occur there will be
significantly affected by interactions. Of course most of
both the experimental and theoretical studies of liquids
are aimed at understanding the effects of interactions. So
liquid-matter scientists are ideally placed to
contribute to attempts to understand the behaviour
in the cytoplasm, in particular to attempts
to understand the differences between {\it in vitro} and {\it in vivo}
behaviour as these are similar to the differences between
ideal gases and dense liquids \cite{minton01,hall03}. 
Finally, in addition to the inevitable interactions
due to the crowded nature of the cytoplasm there are many
interactions that are essential to the function of the cell.
For example, the receptors {\it E.~coli} cells use to
detect nutrient molecules in their environment interact
with each other so as to make their response
cooperative, this has been modelled using a Ising-like model
near a phase transition \cite{duke99,goldman04}. These
receptors are embedded in the cell membrane not in the
cytoplasm but similar interactions may be employed to
produce cooperative phenomena in the cytoplasm.


\section*{Acknowledgements}

I would like to thank the organisers of the 6th Liquid Matter
conference for the invitation to speak and to contribute this
article to the proceedings.
It is also a pleasure to acknowledge invaluable discussions
with J Cuesta and D Frenkel, and
to acknowledge that section \ref{sec_phasesep}
describes work done in collaboration with N Braun, J Cuesta and P Warren.

\end{document}